\documentclass[conference]{IEEEtran}
\IEEEoverridecommandlockouts
\usepackage{cite}
\usepackage{float}
\usepackage{array}
\usepackage{multicol}
\usepackage{multirow}
\usepackage{booktabs, cellspace}
\usepackage{graphicx}
\usepackage{forloop}
\usepackage{amsmath,amssymb,amsfonts}
\usepackage{algorithmic}
\usepackage{graphicx}
\usepackage{longtable}
\usepackage[colorlinks=true,linkcolor=black,urlcolor=black,breaklinks]{hyperref}
\usepackage{fancyhdr}
\usepackage{caption}
\usepackage{tabularx}

\usepackage{url}
\usepackage{textcomp}
\usepackage{xcolor}
\def\BibTeX{{\rm B\kern-.05em{\sc i\kern-.025em b}\kern-.08em
    T\kern-.1667em\lower.7ex\hbox{E}\kern-.125emX}}
\graphicspath{{Figures/}}

\makeatletter
\newcommand*\titleheader[1]{\gdef\@titleheader{#1}}
\AtBeginDocument{%
  \let\st@red@title\@title
  \def\@title{%
    \bgroup\normalfont\large\centering\@titleheader\par\egroup
    \vskip1.5em\st@red@title}
}
\makeatother

\title{A Robust Image Encryption Scheme Based on New
4-D Hyperchaotic System and Elliptic Curve}
\titleheader{\small 2023 4th International Conference on Technological Advances in Electrical Engineering (ICTAEE)\\
University of 20 August 1955, Skikda, Algeria}
\author{\IEEEauthorblockN{Yehia Lalili}
\IEEEauthorblockA{\textit{\small Department of Electrical Engineering}\\
\textit{\small University of Skikda}\\
\small Skikda, Algeria \\
\href{mailto:y.lalili@univ-skikda.dz}{\texttt{y.lalili@univ-skikda.dz}}}\\
\IEEEauthorblockN{Morad Grimes}
\IEEEauthorblockA{\textit{\small Department of Electronics}\\
\textit{\small University of Jijel}\\
\small Jijel, Algeria \\
\href{mailto:grimes_morad@yahoo.fr}{\texttt{grimes\_morad@yahoo.fr}}}
\and
\IEEEauthorblockN{Toufik Bouden}
\IEEEauthorblockA{\textit{\small Department of Automatic} \\
\textit{\small University of Jijel}\\
\small Jijel, Algeria \\
\href{mailto:bouden_toufik@yahoo.com}{\texttt{bouden\_toufik@yahoo.com}}}\\
\IEEEauthorblockN{Abderrazek Lachouri}
\IEEEauthorblockA{\textit{\small Department of Electrical Engineering} \\
\textit{\small University of Skikda}\\
\small Skikda, Algeria \\
\href{mailto:alachouri@yahoo.fr}{\texttt{alachouri@yahoo.fr}}}
}

\begin{document}

\maketitle

\begin{abstract}
In this work, a new \textit{4-D} hyper-chaotic system for image encryption is proposed and its effectiveness is demonstrated by incorporating it into an existing Elliptic Curve Cryptography (\textit{ECC}) mapping scheme. The proposed system is considered simple because it consists of eight terms with two non-linearities. The system exhibits high sensitivity to initial conditions, which makes it suitable for encryption purposes. The two-stage encryption process, involving confusion and diffusion, is employed to protect the confidentiality of digital images. The simulation results demonstrate the effectiveness of the hyper-chaotic system in terms of security and performance when combined with the \textit{ECC} mapping scheme. This approach can be applied in various domains including health-care, military, and entertainment to ensure the robust encryption of digital images.
\end{abstract}
\smallskip

\section{Introduction}
Due to the growing utilization of digital images across various domains including health-care, military, and entertainment, the demand for robust image encryption techniques has been increasing.\\

In recent years, a plethora of image encryption techniques have been presented, such as \textit{DNA} \cite{Zhang, Wang}, quantum computing \cite{Zhou}, compressive sensing \cite{DZhang}, and chaotic-based methods \cite{Lai, Zia, Zolfaghari}. These latter methods possess intrinsic properties like non-periodicity, random behavior, and sensitivity to control parameters and initial conditions. These properties make chaotic-based methods highly effective in encrypting images.\\

Therefore, in this work, a new 4-D hyper-chaotic system for image encryption is proposed and its effectiveness is demonstrated by incorporating it into an existing Elliptic Curve Cryptography (\textit{ECC}) mapping scheme \cite{Soleymani, Lone}. The two-stage encryption process, involving confusion and diffusion, is employed to protect the confidentiality of digital images.\\

The structure of the paper is as follows: In the next section, we delve into the mathematical foundations of the proposed system and image encryption method. In section 3, we elaborate on the design and implementation details of the encryption technique. In section 4, we present the simulation results to demonstrate the effectiveness of the employed method. Finally, in section 5, we conduct a thorough security and performance analysis of the crypto-system.

\section{MATHEMATICAL FOUNDATIONS}
\subsection{Proposed 4-D hyper-chaotic system}
The paper by \textit{Luo}, \textit{Wang} and \textit{Wan} in \cite{LuoWang}, introduces a 3-D chaotic system. Building on this, the technique proposed by \textit{Li et al}. in \cite{LiW} for constructing new 4-D hyper-chaotic systems is applied by adding a linear state feedback controller to the second equation of the 3-D system, resulting in a 4-D autonomous system,
\begin{align}
\begin{split}
\dot{x} &= a(y-x),\\
\dot{y} &= -e_1 xz + cy + kw,\\
\dot{z} &= -b + e_2 y^2,\\
\dot{w} &= - my,
\end{split}
\end{align}
in which $[x,y,z,w]^T$ is state vector. The parameters $a,\, b,\, c,\, e_1,\, e_2,\, k$ and $m$ are positive constants.

The proposed system is considered simple because it consists of eight terms with two non-linearities. When the parameters of system (1) are taken as:
\begin{equation}
\begin{gathered}
a=10, \, b=3, \, c=2.5, \, e_1=12,\\
e_2=0.1, \, m=2, \, k=2.
\end{gathered}
\end{equation}
and the initial conditions as:
\begin{equation}
x_0=1, \, y_0=1, \, z_0=1, \, w_0=1.
\end{equation}
System (1) has a hidden hyper-chaotic attractor, with phase portraits as depicted in Fig. 1.

\subsection{Lyapunov exponents}
When we take the constants as in (2) and the initial states as in (3), the \textit{Lyapunov} exponents of the model (1) can be calculated using the \textit{Wolf} algorithm \cite{Wolf}.
\begin{equation}
LE_1=0.971, \, LE_2=0.102, \, LE_3=0, \, LE_4=-8.819.
\end{equation}
As shown in Fig. 2, there are two positive \textit{Lyapunov} exponents. Hence, the proposed system (1) is hyper-chaotic.
\subsection{Equilibrium points}
The equilibrium points can be determined by solving the given system of algebraic equations:
\begin{align}
a(y-x) &= 0,\\
-e_1 xz + cy + kw &= 0,\\
-b + e_2 y^2 &= 0,\\
- my &= 0.
\end{align}
Upon evaluating (8), we obtain the result $y = 0$ . By substituting this value into (7), we deduce that $b$ must be equal to $0$ . However, this conflicts with the given condition that $b > 0$ . Hence, it can be concluded that the proposed system does not possess any equilibrium points. Consequently, the system described by (1) falls into the class of systems exhibiting hidden attractors.

\begin{figure}[H]
\centering
\includegraphics[scale=1]{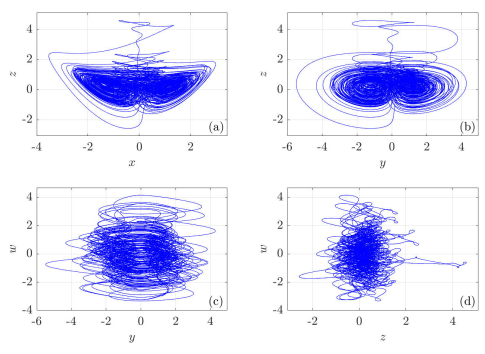}
\caption{Hyper-chaotic attractor of system (1): (a) $x-z$ plane, (b) $y-z$ plane, (c) $y-w$ plane and (d) $z-w$ plane.}
\label{fig:1}
\end{figure}
\subsection{Sensitivity to initial conditions}
Fig. 3 illustrates the chaotic behavior that results from small variations in initial values ($\pm 10^{-15}$) due to the high sensitivity of the hyper-chaotic system (1) to initial conditions.

\subsection{Elliptic curve cryptography}
\begin{itemize}
\item[-] An elliptic curve defined by:
\begin{equation}
E_P\,(a,b): \, y^2=x^3+ax+b \, mod \, p,
\end{equation}
where $a$ and $b$ are integers and $p$ is a large prime number, and it must also satisfy the condition:
\begin{equation}
4a^3+27b^2\neq 0\, mod \, p.
\end{equation}

\item[-] An affine point $G(X,Y)$ that lies on the curve.
\item[-] Selection of private keys $x$, $y$, by \textit{Alice} and \textit{Bob} respectively, and the computation of public keys:
\begin{equation}
P_A=xG,\; P_B=yG.
\end{equation}
\item[-] Encryption of the message by \textit{Alice} using a random integer $k$ and \textit{Bob's} public key $P_B$ to create:
\begin{equation}
P_C = \left[ \left( kG\right), \left( P_M+kP_B\right) \right].
\end{equation}
\item[-] Decryption of $P_C$ by \textit{Bob} using his private key $y$ to retrieve the original message $P_M$ by performing:
\begin{equation}
P_M=(P_M+kP_B) - [ykG].
\end{equation}
\item[-] Addition operation for two points $P$ and $Q$ over an elliptic group, $P+Q=(X_3,Y_3)$ is given by:
\begin{equation}
X_3=\lambda^2 - X_P - X_Q \, mod\, p,
\end{equation}
\begin{equation}
Y_3=\lambda (X_P - X_3) - Y_P \, mod\, p,
\end{equation}
where:
\begin{equation}
\lambda = \left\{\begin{matrix}
\frac{Y_Q-Y_P}{X_Q-X_P}\, mod \, p,\; \text{if} \; P\neq Q\; \text{(point addition)}. \\
 \\ 
\frac{3X^2_P+a}{2Y_P}\, mod \, p, \; \text{if} \; P=Q\; \text{(point doubling)}.
\end{matrix}\right.
\end{equation}
\end{itemize}

\begin{figure}[h]
\centering
\includegraphics[scale=1]{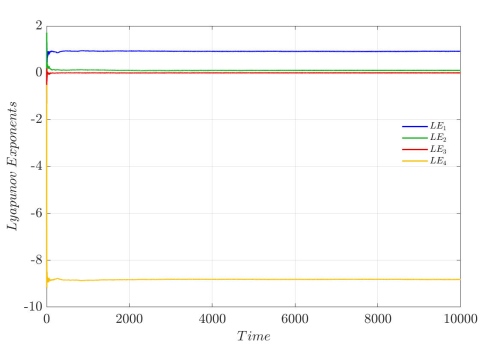}
\caption{\textit{Lyapunov} exponents of the new system (1).}
\label{fig:2}
\end{figure}

\begin{figure}[h]
\centering
\includegraphics[scale=1]{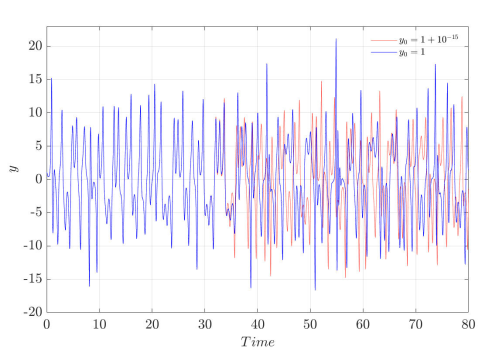}
\caption{Time series of the $y$ variable.}
\label{fig:3}
\end{figure}
\section{CRYPTO-SYSTEM APPROACH}
Image encryption based on hyper-chaos involves modifying the arrangement and pixel values within an image using two sequential stages: confusion and diffusion. The encryption process is illustrated in Fig. 4, while the decryption process is depicted in Fig. 5, showcasing the proposed flowcharts for each stage.

\section{SIMULATION RESULTS}
To validate the effectiveness of the suggested crypto-system, we used \textit{MATLAB 2022} on a personal computer with an \textit{11th Gen Intel(R) Core(TM) i5-1135G7 @2.40GHz 2.42 GHz} processor, \textit{16 GB} of \textit{RAM}, and a \textit{Windows 10} operating system. The test image used in the simulation was the Peppers image $256 \times 256$.\\

The system parameters of the hyper-chaotic system and
the \textit{ECC} keys used in the simulation are listed in Table. 1. The results of the simulation are depicted in Fig. 6.

\subsection{Key sensitivity}
To assess the sensitivity of the encryption keys in the suggested crypto-system, an experiment was conducted using different initial conditions of the 4-D hyper-chaotic system. The original initial conditions of $(1,1,1,1)$ were modified to $(1,1+10^{-15} ,1,1)$ which represents a small change in the key values. The encryption and decryption process was performed using both sets of initial conditions and the results were compared.
\begin{table}[H]
\centering
\caption{Experiment parameters.}
\label{tab:1}
\resizebox{0.5\textwidth}{!}{%
\begin{tabular}{@{}lc@{}}
\toprule
\multicolumn{1}{c}{Item} &
  Value \\ \midrule
\begin{tabular}[c]{@{}l@{}}Parameters of  the hyper-chaotic system (1)\\ $ $ \\ $ $\end{tabular} &
  \begin{tabular}[c]{@{}c@{}}$a=10$, $b=3$, $c=2.5$, $e_1=12$, $e_2=0.1$,\\ $m=2$, $k=2$, ($x_0$, $y_0$, $z_0$, $w_0$) = ($1$, $1$, $1$, $1$)\\ $ $\end{tabular} \\
\begin{tabular}[c]{@{}l@{}}\textit{ECC} parameters\\ $ $\end{tabular} &
  \begin{tabular}[c]{@{}c@{}}$a=5376$, $b=2438$, $p=123457$, $y=36548$,\\ $k=23412$, $P_B=(30402,\, 35513)$, $G=(2225,\, 75856)$\end{tabular} \\ \bottomrule
\end{tabular}%
}
\end{table}

\begin{figure}[H]
\centering
\includegraphics[scale=1]{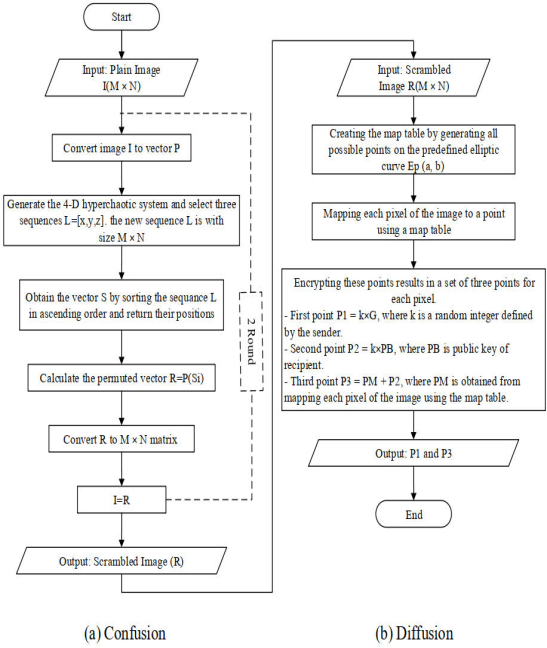}
\caption{Flowchart diagram for the encryption process.}
\label{fig:4}
\end{figure}

\begin{figure}[H]
\centering
\includegraphics[scale=0.95]{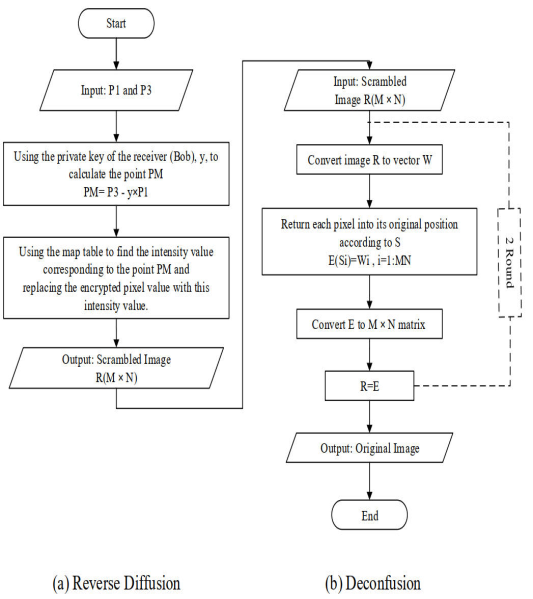}
\caption{Flowchart diagram for the decryption process.}
\label{fig:5}
\end{figure}
 The experimental findings illustrate that a minor alteration in the initial conditions of the hyper-chaotic system exerts a substantial influence on the security of the implemented crypto-system. Upon utilizing modified initial conditions, the decryption process proved incapable of recovering the original image, underscoring the pronounced susceptibility of the implemented crypto-system to the key values employed during the encryption procedure. The experiment results were visualized in Fig. 7.

\subsection{Key space}
The key size plays an important role in the security of the crypto-system. The larger the key size, the more difficult it is for an attacker to perform a Brute Force attack. The implemented crypto-system uses \textit{ECC} which provides an exponentially difficult Elliptic Curve Discrete Logarithmic Problem (\textit{ECDLP}) with respect to the key size \cite{Lone}. The key size used in this implementation is quite large, making it even more difficult for an attacker to successfully perform a Brute Force attack.

\subsection{Histogram}
The crypto-system was evaluated for its ability to resist statistical attacks by comparing the distribution of pixel values in the original and encrypted images. The results of the analysis, shown in Fig. 6, demonstrate that the encrypted images have a uniform and distinct distribution of pixel values, making it difficult for an attacker to determine any information about the original image. The implemented crypto-system was able to achieve this level of security by carefully managing the key values used in the encryption process, as discussed in previous sections. Overall, the results of this analysis demonstrate the robustness and effectiveness of the crypto-system in protecting the confidentiality of digital images.

\subsection{Correlation analysis}
In this study, we conducted a correlation analysis on the encrypted image using the Pearson correlation coefficient to evaluate the performance of the crypto-system. The correlation coefficients between the original and the encrypted image were calculated using the formula:
\begin{equation}
r_{x,y}=\frac{E\left( \left( x-E(x)\right) \cdot \left( y- E(y)\right) \right) }{\sqrt{D(x) \cdot D(y)}},
\end{equation}
\begin{equation}
E(x)= \frac{1}{N} \sum_{i=1}^{N}x_i,
\end{equation}
\begin{equation}
D(x)=\frac{1}{N} \sum_{i=1}^{N} \left( x_i - E(x)\right)^2,
\end{equation}
where $E(x)$ and $D(x)$ represent the mean and variance of the variable $x$ respectively. The results of the correlation analysis are presented in Fig. 8, which illustrates the distribution of adjacent pixels of the plain image ``peppers" and its cipher images, and in Table. 2, which shows the Correlation coefficients of the plain and cipher images.
\begin{figure}[H]
\centering
\includegraphics[scale=1]{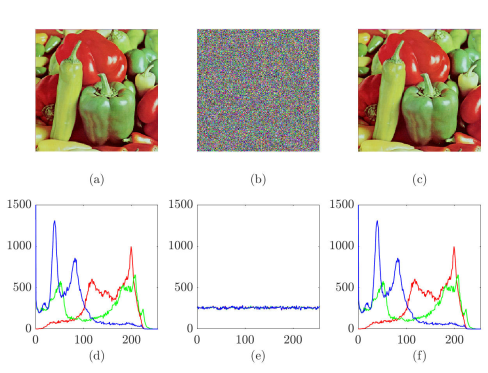}
\caption{The result of the proposed crypto-system (a) Original ``Peppers" image, (b) Encrypted ``Peppers" image, (c) Decrypted image, (d) Histogram of (a), (e) Histogram of (b), (f) Histogram of (c).}
\label{fig:6}
\end{figure}

 The results show that the correlation coefficient $r_{x,y}$ of the encrypted image is close to zero, indicating that the implemented crypto-system effectively eliminates the correlation between adjacent pixels and improves the security against statistical attacks.

\subsection{Differential attack}
In this differential attack, the attacker attempts to decrypt the encrypted image without the use of a key by identifying the relationship between the original and encrypted images. This is a significant concern for image encryption algorithms as small changes in the original image should result in significant changes in the encrypted image, making it more difficult for the attacker to decrypt the image. \\

To evaluate the robustness of the encryption algorithm against this type of attack, we used the Number of Pixels Change Rate (\textit{NPCR}) and Unified Average Changing Intensity (\textit{UACI}) as metrics. These metrics are commonly used to measure the sensitivity of an encryption algorithm to small changes in the original image and demonstrate its ability to resist a differential attack. 

The \textit{NPCR} is a measure of how many pixels are different between two cipher images obtained from the same encryption
\begin{table}[H]
\centering
\caption{Correlation coefficients of the plain and cipher image.}
\label{tab:2}
\resizebox{0.5\textwidth}{!}{%
\begin{tabular}{@{}lcccccc@{}}
\toprule
\multicolumn{1}{c}{Direction} & \multicolumn{6}{c}{Correlation coefficients}                       \\ \midrule
                              & \multicolumn{3}{c}{Plain image} & \multicolumn{3}{c}{Cipher Image} \\ \cmidrule(l){2-7} 
                              & Red       & Green    & Blue     & Red      & Green     & Blue      \\ \midrule
Horizontal                    & 09424     & 0.9559   & 0.9319   & 0.0034   & -0.0013   & -0.007    \\
Vertical                      & 0.9464    & 0.9617   & 0.9406   & 0.0032   & -0.0063   & -0.0002   \\
Diagonal                      & 0.9094    & 0.9285   & 0.8981   & 0.0002   & -0.0006   & 0.0009    \\ \bottomrule
\end{tabular}%
}
\end{table}
\noindent key, and \textit{UACI} measures the average intensity of differences between the two cipher images. They are defined as follows:

\begin{equation}
NPCR_{R,G,B}=\frac{\sum_{i,j}D_{R,G,B}(i,j)}{W \times H}\times 100 \%.
\end{equation}

\begin{equation}
 \begin{matrix}
UACI_{R,G,B} =\frac{1}{W \times H} \times\\
\\
\left[ \sum_{i,j}\frac{\mid C_{2_{R,G,B}}(i,j)-C_{1_{R,G,B}}(i,j)\mid}{255}\right] \times 100\%.
\end{matrix}
\end{equation}
with:
\begin{equation}
D_{R,G,B}(i,j)= 
\left\{\begin{matrix}
0, \;\; C_{2_{R,G,B}}(i,j) = C_{1_{R,G,B}}(i,j).\\
0, \;\; C_{2_{R,G,B}}(i,j) \neq C_{1_{R,G,B}}(i,j).
\end{matrix}\right.
\end{equation}
The symbol $C_2$ refers to the chipper image that encrypted from the original image by changing only one pixel, while $C_1$ refers to the chipper image encrypted from the same plain image. \textit{NPCR} and \textit{UACI} values for an ideal image with size $256\times 256$ should be larger then $99.5693\%$ and in the range $33.2824\%$, $33.6447\%$, respectively. In our experiment, we used the Peppers image and changed the pixel as follows:
\begin{equation}
\begin{small}
\begin{gathered}
Peppers(100,100,1)=140\Rightarrow Peppers(100,100,1)=141\\
Peppers(100,100,2)=23\Rightarrow Peppers(100,100,2)=24 \\
Peppers(100,100,3)=28\Rightarrow Peppers(100,100,3)=29
\end{gathered}
\end{small}
\end{equation}
The results of our \textit{NPCR} and \textit{UACI} calculations are presented in Table. 3. As shown in the table, our crypto-system demonstrates excellent robustness against the differential attack with \textit{NPCR} and \textit{UACI} values that are close to the ideal values.

\subsection{Data loss}
A robust image encryption system should be able to withstand data loss during transmission. In order to test the crypto-system resistance to data loss, two simulations were conducted. The first simulation involved cutting a $50 \times 50$ section of the encrypted Peppers image and
attempting to decrypt the remaining data. The second simulation involved cutting a $100\times 100$ section of the same encrypted Peppers image and attempting to decrypt the remaining data. The results, shown in Fig. 9, indicate that even with significant data loss, the decrypted image still retains a majority of the original information.

\begin{table}[H]
\centering
\caption{Results of average \textit{NPCR} and \textit{UACI} values.}
\label{tab:3}
\resizebox{0.5\textwidth}{!}{%
\begin{tabular}{@{}cccccc@{}}
\toprule
\multicolumn{3}{c|}{$NPCR$}             & \multicolumn{3}{c}{$UACI$}            \\ \midrule
Red         & Green       & Blue        & Red       & Green       & Blue        \\ \midrule
$99.5911\%$ & $99.6277\%$ & $99.5209\%$ & $33.3946$ & $33.4467\%$ & $33.4536\%$ \\ \bottomrule
\end{tabular}%
}
\end{table}

\begin{figure}[H]
\centering
\includegraphics[scale=0.8]{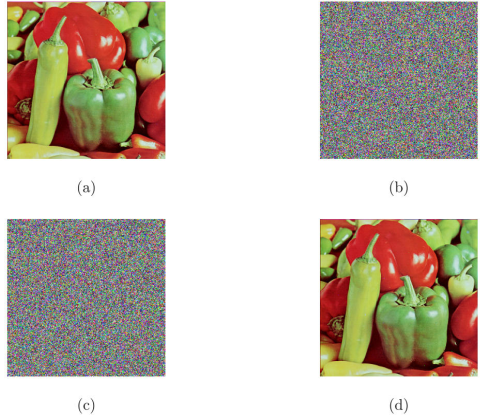}
\caption{Key sensitivity: (a) original ``Papers", (b) encrypted ``Papers" with the original initial conditions, (c) decrypted ``Papers" with the modified key, and (d) decrypted ``Papers" with the original key.}
\label{fig:7}
\end{figure}

\begin{figure}[H]
\centering
\includegraphics[scale=0.8]{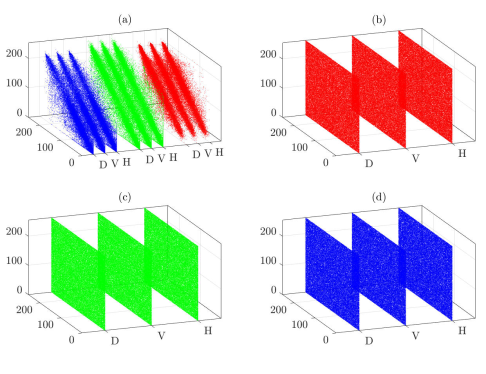}
\caption{Correlation distributions of adjacent pixels in the horizontal, vertical, diagonal directions: (a) distribution of original color image Peppers, (b–d) distributions of red, green, and blue components of encryption image Peppers, respectively.}
\label{fig:8}
\end{figure}
\begin{figure}[H]
\centering
\includegraphics[scale=0.8]{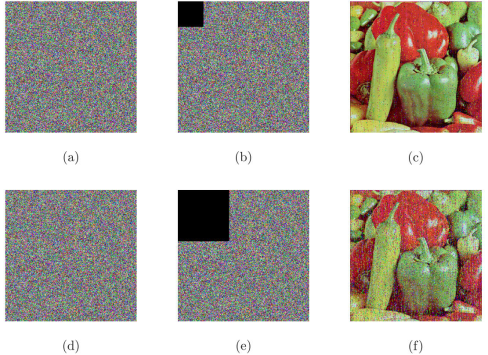}
\caption{Data loss analysis: (a \& d) Encrypted image, (b) Encrypted image with $50\times 50$ data cut, (c) Decrypted image of (b), (e) Encrypted image with $100 \times 100$ data cut, (f) Decrypted image of (e).}
\label{fig:9}
\end{figure}

\section{Conclusion}
In conclusion, the study presented a novel 4-D hyper-chaotic system and introduced a crypto-system that utilizes this system for image encryption, specifically in the confusion stage. The employed crypto-system employs a two stage encryption process, incorporating confusion and diffusion, to ensure the confidentiality of digital images. \\

Several experiments were conducted to analyze the security and performance of the devised crypto-system. The outcomes of these experiments demonstrate the high sensitivity of the utilized crypto-system to key values, its expansive key space, its resilience against statistical attacks, and its substantial level of randomness. \\

Specifically, the results of the key sensitivity experiment highlighted the significant impact on the crypto-system's security when even slight modifications are made to the initial conditions of the hyper-chaotic system during the confusion stage. This emphasizes the utmost importance of securely managing the key values during the encryption process. Overall, the study showcases a robust and effective approach to safeguarding the confidentiality of digital images through the integration of the hyper-chaotic system in the confusion stage.


\begin{thebibliography}{20}
\bibitem{Zhang}
Y. Zhang, "The image encryption algorithm based on chaos and DNA computing," Multimedia Tools and Applications, vol. 77, no. 16, pp. 21589-21615, 2018.

\bibitem{Wang}
S. Wang, Q. Peng, and B. Du, "Chaotic color image encryption based on 4-D chaotic maps and DNA sequence," Optics \& Laser Technology, vol. 148, pp. 107753, 2022.

\bibitem{Zhou}
N. Zhou, Y. Hu, L. Gong, and G. Li, "Quantum image encryption scheme with iterative generalized Arnold transforms and quantum image cycle shift operations," Quantum Information Processing, vol. 16, pp. 1-23, 2017.

\bibitem{DZhang}
D. Zhang, X. Liao, B. Yang, and Y. Zhang, "A fast and efficient approach to color-image encryption based on compressive sensing and fractional Fourier transform," Multimedia Tools and Applications, vol. 77, pp. 2191-2208, 2018.

\bibitem{Lai}
Q. Lai, H. Zhang, P. D. K. Kuate, G. Xu, and X.-W. Zhao, "Analysis and implementation of no-equilibrium chaotic system with application in image encryption," Applied Intelligence, vol. 52, no. 10, pp. 11448-11471, 2022.

\bibitem{Zia}
U. Zia, M. McCartney, B. Scotney, J. Martinez, M. AbuTair, J. Memon, and A. Sajjad, "Survey on image encryption techniques using chaotic maps in spatial, transform and spatiotemporal domains," International Journal of Information Security, vol. 21, no. 4, pp. 917- 935, 2022.

\bibitem{Zolfaghari}
B. Zolfaghari and T. Koshiba, "Chaotic Image Encryption: State-ofthe-Art, Ecosystem, and Future Roadmap," Applied System Innovation, vol. 5, no. 3, pp. 57, 2022.

\bibitem{Soleymani}
A. Soleymani, M. J. Nordin, A. N. Hoshyar, Z. M. Ali, and E. Sundararajan, "An image encryption scheme based on elliptic curve and a novel mapping method," International Journal of Digital Content Technology and its Applications, vol. 7, no. 13, pp. 85, 2013.

\bibitem{Lone}
P. N. Lone, D. Singh, V. Stoffová, D. C. Mishra, U. H. Mir, and N. Kumar, "Cryptanalysis and Improved Image Encryption Scheme Using Elliptic Curve and Affine Hill Cipher," Mathematics, vol. 10, no. 20, pp. 3878, 2022.

\bibitem{LuoWang}
X. Luo, C. Wang, and Z. Wan, "Grid multi-wing butterfly chaotic attractors generated from a new 3-D quadratic autonomous system," Nonlinear Analysis: Modelling and Control, vol. 19, no. 2, pp. 272- 285, 2014.

\bibitem{LiW}
Y. Li, W. K. Tang, and G. Chen, "Generating hyperchaos via state feedback control," International Journal of Bifurcation and Chaos, vol. 15, no. 10, pp. 3367-3375, 2005.

\bibitem{Wolf}
A. Wolf, J. B. Swift, H. L. Swinney, and J. A. Vastano, "Determining Lyapunov exponents from a time series," Physica D: Nonlinear Phenomena, vol. 16, no. 3, pp. 285-317, 1985.

\end{thebibliography}
\end{document}